\PassOptionsToPackage{unicode}{hyperref}
\PassOptionsToPackage{hyphens}{url}
\PassOptionsToPackage{dvipsnames,svgnames,x11names}{xcolor}
\documentclass[
  12pt]{article}

\usepackage{amsmath,amssymb}
\usepackage{iftex}
\ifPDFTeX
  \usepackage[T1]{fontenc}
  \usepackage[utf8]{inputenc}
  \usepackage{textcomp} 
\else 
  \usepackage{unicode-math}
  \defaultfontfeatures{Scale=MatchLowercase}
  \defaultfontfeatures[\rmfamily]{Ligatures=TeX,Scale=1}
\fi
\usepackage{lmodern}
\ifPDFTeX\else  
\fi
\IfFileExists{upquote.sty}{\usepackage{upquote}}{}
\IfFileExists{microtype.sty}{
  \usepackage[]{microtype}
  \UseMicrotypeSet[protrusion]{basicmath} 
}{}
\makeatletter
\@ifundefined{KOMAClassName}{
  \IfFileExists{parskip.sty}{%
    \usepackage{parskip}
  }{
    \setlength{\parindent}{0pt}
    \setlength{\parskip}{6pt plus 2pt minus 1pt}}
}{
  \KOMAoptions{parskip=half}}
\makeatother
\usepackage{xcolor}
\makeatletter
\ifx\paragraph\undefined\else
  \let\oldparagraph\paragraph
  \renewcommand{\paragraph}{
    \@ifstar
      \xxxParagraphStar
      \xxxParagraphNoStar
  }
  \newcommand{\xxxParagraphStar}[1]{\oldparagraph*{#1}\mbox{}}
  \newcommand{\xxxParagraphNoStar}[1]{\oldparagraph{#1}\mbox{}}
\fi
\ifx\subparagraph\undefined\else
  \let\oldsubparagraph\subparagraph
  \renewcommand{\subparagraph}{
    \@ifstar
      \xxxSubParagraphStar
      \xxxSubParagraphNoStar
  }
  \newcommand{\xxxSubParagraphStar}[1]{\oldsubparagraph*{#1}\mbox{}}
  \newcommand{\xxxSubParagraphNoStar}[1]{\oldsubparagraph{#1}\mbox{}}
\fi
\makeatother

\usepackage{longtable,booktabs,array}
\usepackage{calc} 
\usepackage{etoolbox}
\makeatletter
\patchcmd\longtable{\par}{\if@noskipsec\mbox{}\fi\par}{}{}
\makeatother
\IfFileExists{footnotehyper.sty}{\usepackage{footnotehyper}}{\usepackage{footnote}}
\makesavenoteenv{longtable}
\usepackage{graphicx}
\makeatletter
\def\maxwidth{\ifdim\Gin@nat@width>\linewidth\linewidth\else\Gin@nat@width\fi}
\def\maxheight{\ifdim\Gin@nat@height>\textheight\textheight\else\Gin@nat@height\fi}
\makeatother
\setkeys{Gin}{width=\maxwidth,height=\maxheight,keepaspectratio}
\makeatletter
\def\fps@figure{htbp}
\makeatother

\makeatletter
\@ifpackageloaded{caption}{}{\usepackage{caption}}
\AtBeginDocument{%
\ifdefined\contentsname
  \renewcommand*\contentsname{Table of contents}
\else
  \newcommand\contentsname{Table of contents}
\fi
\ifdefined\listfigurename
  \renewcommand*\listfigurename{List of Figures}
\else
  \newcommand\listfigurename{List of Figures}
\fi
\ifdefined\listtablename
  \renewcommand*\listtablename{List of Tables}
\else
  \newcommand\listtablename{List of Tables}
\fi
\ifdefined\figurename
  \renewcommand*\figurename{Figure}
\else
  \newcommand\figurename{Figure}
\fi
\ifdefined\tablename
  \renewcommand*\tablename{Table}
\else
  \newcommand\tablename{Table}
\fi
}
\@ifpackageloaded{float}{}{\usepackage{float}}
\floatstyle{ruled}
\@ifundefined{c@chapter}{\newfloat{codelisting}{h}{lop}}{\newfloat{codelisting}{h}{lop}[chapter]}
\floatname{codelisting}{Listing}

\makeatother
\makeatletter
\@ifpackageloaded{caption}{}{\usepackage{caption}}
\@ifpackageloaded{subcaption}{}{\usepackage{subcaption}}
\makeatother

\ifLuaTeX
  \usepackage{selnolig}  
\fi
\usepackage[]{natbib}
\usepackage{bookmark}

\usepackage{lineno}

\IfFileExists{xurl.sty}{\usepackage{xurl}}{} 
\hypersetup{
  pdftitle={Title},
  pdfauthor={Author 1; Author 2},
  pdfkeywords={3 to 6 keywords, that do not appear in the title},
  colorlinks=true,
  linkcolor={blue},
  filecolor={Maroon},
  citecolor={Blue},
  urlcolor={Blue},
  pdfcreator={LaTeX via pandoc}}

\newcommand{\anon}{1}

\usepackage{algorithm}
\usepackage{algpseudocode}
\usepackage{booktabs}

\newcommand{\bs}{\textbf{s}}
\newcommand{\bu}{\textbf{u}}

\begin{document}

\def\spacingset#1{\renewcommand{\baselinestretch}%
{#1}\small\normalsize} \spacingset{1}


\if1\anon
{
  \title{\bf Efficient Bayesian Inference for Spatial Point Patterns Using the Palm Likelihood}
  \author{Kevin M. Collins\thanks{This work is supported in part by U.S. Geological Survey Northeast and Southeast Climate Adaptation Science Center Grant No. G22AC00597-01.}\hspace{.2cm}\\
    Department of Statistics, North Carolina State University\\
    and \\
    Erin M. Schliep \\
    Department of Statistics, North Carolina State University}
  \maketitle
} \fi

\if0\anon
{
  \bigskip
  \bigskip
  \bigskip
  \begin{center}
    {\LARGE\bf Efficient Bayesian Inference for Spatial Point Patterns Using the Palm Likelihood}
\end{center}
  \medskip
} \fi

\bigskip
\begin{abstract}
Bayesian inference for spatial point patterns is often hindered computationally by intractable likelihoods. In the frequentist literature, estimating equations utilizing pseudolikelihoods have long been used for parameter estimation. One such pseudolikelihood based on the process of differences is known as the Palm likelihood. Utilizing notions of Bayesian composite likelihoods and generalized Bayesian inference, we develop a framework for the use of Palm likelihoods in a Bayesian context with implementation using the probabilistic programming language, Stan. Naive implementation of the Palm likelihood results in posterior undercoverage of model parameters. We propose two approaches to remedy this issue and calibrate the resulting posterior. Numerical simulations illustrate both the efficacy of the method in terms of statistical properties and the superiority in terms of computational efficiency when compared to classical Bayesian inference. The method is then applied to the popular \textit{Beilschmiedia pendula Lauraceae} dataset.
\end{abstract}

\noindent%
\vfill

\newpage
\spacingset{1.8} 

\section{Introduction}

Spatial point pattern data arises in research areas such as ecology \citep{waagepetersen2016analysis, velazquez2016evaluation, perry2006comparison, ben2021spatial}, biology \citep{weston2012analysis, burguet2014statistical}, criminology \citep{shirota2017space}, seismology \citep{d2023locally}, and public health \citep{diggle2013spatial}, and have been studied extensively by statisticians and probabilists. A general theoretical treatment of point processes can be found in \cite{moller2003statistical}, with emphasis on spatial point processes presented in \cite{diggle2013statistical} and \cite{baddeley2016spatial}. Statistical estimation and inferential methods for spatial point patterns are often complex and difficult to implement, which limits their utility. In particular, many of these methods are computationally intractable, requiring advanced algorithms for model fitting. 

For point patterns, the likelihood itself is often intractable. Take, for example, the popular log-Gaussian Cox process (LGCP) \citep{moller1998log}. The likelihood requires integration over an infinite-dimensional random variable due to its doubly stochastic nature, requiring some form of approximation. Further, the assumption of a latent Gaussian process requires the inversion of a covariance matrix, which scales cubically in complexity with the dimension of the discretized process. One approach to subvert this intractability is through the optimization of an objective function simpler than the likelihood. Minimum contrast estimation \citep{baddeley2016spatial} is a popular approach in the spirit of the method of moments applied to point processes. Letting $\boldsymbol\theta$ be the parameter of interest and $u$ be the interpoint distance, an estimate $\boldsymbol{\hat\theta}$ is obtained by minimizing the discrepancy measure
\begin{equation}\label{mce}
    D(\boldsymbol\theta)=\int_0^r ||\hat J(u)^c-J(u;\boldsymbol\theta)^c||^p du
\end{equation}
for preselected constants $r,c$, and $p$ where $J(\cdot)$ is a summary function, which has a closed form, such as the pair-correlation function or the K-function, and $\hat J(u)$ can be  estimated nonparameterically. The primary advantage is that it is computationally simple to obtain parameter estimates; however, the need to select the three tuning parameters is a drawback.

Likelihood-based inference is often preferred by practitioners, but this can be costly. Computationally efficient likelihood-based methods have been developed based on the idea of composite likelihoods \citep{lindsay11988composite,varin2011overview} or pseudolikelhoods \citep{besag1974spatial} in a number of different settings. One such example is semivariogram estimation in geostatistical models \citep{curriero1999composite, schliep2023correcting}, which was extended to a spatiotemporal setting \citep{bevilacqua2012estimating}.

In the context of point patterns, \cite{baddeley2000practical} first developed a pseudolikelihood approach for Markov point processes. For cluster and Cox processes, \cite{guan2006composite} proposed a composite likelihood based on second-order intensities, whereas the Palm likelihood based on palm intensities was developed by \cite{ogata1991maximum} and \cite{tanaka2008parameter}. Each of these "likelihoods" is available in closed form, which dramatically improves computational performance. In the cases for point patterns, practitioners have only ever considered pairs of points when constructing the composite likelihood, leading to simplistic expressions. The drawback is that composite likelihoods result in a loss of information when compared to the full likelihood. 

To date, these pseudolikelihood methods have been developed within the frequentist framework independent of the Bayesian literature. 
Classical Bayesian inference is an attractive approach to modeling point processes as the inherent hierarchical structure allows for the modeling of latent processes and enables full posterior inference with uncertainty propagation. 
Inference for Bayesian hierarchical models for point processes offer their own unique challenges, with computational difficulties often exacerbated by the iterative nature of Markov chain Monte Carlo (MCMC) sampling. 
Whereas the aforementioned frequentist approaches work for a general class of point processes, unique approaches for inference are required for specific classes of point processes. For example, when fitting an LGCP, one has to employ an approach such as the Metropolis Adjusted Langevin Algorithm \citep{moller1998log}, Hamiltonian Monte Carlo \citep{betancourt2017conceptual, teng2017bayesian} or Elliptical slice sampling \citep{murray2010elliptical} in order to successfully explore the high-dimensional space of the latent Gaussian process. There has also been a push to develop deterministic computational strategies for LGCPs including INLA \citep{rue2017bayesian} and variational Bayes \citep{teng2017bayesian}. As for inhibition and clustering processes, there is significantly less literature on Bayesian computation. \cite{gelfand2018bayesian} advocate for the use of Approximate Bayesian Computation (ABC) for these classes of process. For Neyman-Scott processes, one has to take an approach such as the birth-death-move \citep{kopecky2016bayesian} in order to successfully sample from the parameter space. 

Frequentists have adapted to the computational complexity of these models with likelihood approximations. Recently, a body of literature has been developing around replacing the likelihood in Bayesian computation with a simpler objective function. For example, \cite{pauli2011bayesian} proposed the use of composite likelihoods in the Bayesian setting, and \cite{ribatet2012bayesian} used the approach in a geostatistical context for extremes analysis. The notion of replacing the likelihood with another function that links the data to the parameters of interest (i.e. loss functions) has been studied in the literature as generalized Bayesian inference \citep{bissiri2016general}, Gibbs posteriors \citep{martin2022direct}, or Laplace-type estimators \citep{chernozhukov2003mcmc}. Importantly, \cite{chernozhukov2003mcmc} and \cite{kleijn2012bernstein} develop relevant Bernstein-von Mises results showing that the resulting posterior accumulates mass asymptotically around the estimator obtained from optimizing the objective function. Unfortunately, these results also show that resulting credible intervals provide inadequate coverage in terms of frequentist error rates. This requires calibration of the posterior \citep{wu2023comparison,ribatet2012bayesian,shaby2014open,pauli2011bayesian} to control the error rate.

Our contribution in this paper is to extend the use of Palm likelihoods into a Bayesian framework for parameter inference in spatial point processes. We develop two different approaches to calibrating the resulting posteriors, which present unique challenges due to the nature of point patterns. Further, we develop an empirical Bayes approach based on the overall intensity of points that mimics the behavior of similar frequentist estimators. Throughout, we illustrate the efficacy of this generalized Bayesian approach in estimation, as well as the computational simplicity of our approach. The Palm likelihood enables direct sampling from the parameter space regardless of the class of processes. Additionally, in contrast to INLA or variational Bayes, our approach is viable for multiple classes of point processes exhibiting different behaviors, not limited to those with a latent Gaussian process. The flexibility and simplicity of the Palm intensity expressions allows for easy model comparisons across a  suite of different processes.

In Section 2, we establish the necessary background on spatial point processes and the Palm likelihood. In Section 3, we discuss the implementation of Palm likelihoods in a Bayesian framework and develop algorithms for calibrating posterior uncertainty. Then, in Sections 4 and 5 we demonstrate the computational efficiency and inferential performance of our approach on simulated point patterns and a real data set, respectively. We conclude with a discussion of the work and future research in Section 6.

\section{Palm Likelihood for Spatial Point Processes}

\subsection{Background}\label{point process background}

Let $\mathcal{S}=\{\textbf{s}_1,\dots,\textbf{s}_n\}$ be a spatial point pattern realized on the compact domain $\mathcal D\subset \mathbb R^2$ from a process indexed by parameter $\boldsymbol\theta$. For a region $B \subset \mathcal D$, let $N(B)$ denote the number of points of $\mathcal{S}$ in $B$. 
Let $d\textbf{s}$ denote the infinitesimal ball around location $\textbf{s}$ and $|d\textbf{s}|$ its area. The first-order intensity of the point pattern is defined as
\begin{equation}\label{first order intensity}
\lambda(\textbf{s}) = \lim_{|d\textbf{s}|\rightarrow0}\frac{E[N(d\textbf{s})]}{|d\textbf{s}|},
\end{equation}
where $\lambda(\textbf{s})|d\textbf{s}|$ is the probability of observing an event in $d\bs$. Further, the probability that two points are observed at locations $\textbf{s}_i$ and $\textbf{s}_j$ is given by $\lambda_2(\textbf{s}_i,\textbf{s}_j)|d\textbf{s}_i||d\textbf{s}_j|$ where the second-order intensity is defined as
\begin{equation}
\lambda_2(\textbf{s}_i,\textbf{s}_j) = \lim_{|d\textbf{s}_i|,|d\textbf{s}_j|\rightarrow 0} \frac{E[N(d\textbf{s}_i)N(d\textbf{s}_j)]}{|d\textbf{s}_i||d\textbf{s}_j|}.
\end{equation}
The conditional expression of the intensity of a point process, known as the Palm intensity \citep{coeurjolly2017tutorial}, is written
\begin{equation}
\lambda_p(\textbf{s}_i|\textbf{s}_j)=\lambda_2(\textbf{s}_i,\textbf{s}_j)/\lambda(\bs_j).
\end{equation}
The Palm intensity times the area, $\lambda_p(\textbf{s}_i|\textbf{s}_j)|d\textbf{s}_i|$, can be interpreted as the probability of observing an event at location $\textbf{s}_i$ given we observed an event at location $\textbf{s}_j$.
Conveniently, an alternative specification of the Palm intensity is 
\begin{equation}
\lambda_p(\textbf{s}_i|\textbf{s}_j)=\lambda(\textbf{s}_i)g(\textbf{s}_i,\textbf{s}_j),
\end{equation}
which is the product of the first-order intensity, $\lambda(\mathbf{s}_i)$, and the pair correlation function, $g(\bs_i,\bs_j)$. An introduction to the Palm distribution for statisticians can be found in \cite{coeurjolly2017tutorial}, which includes intuition as well as the measure-theoretic foundations for analyzing its properties. 

Likelihood-based inference is an attractive and principled option for statistical analysis. However, many likelihood functions for point processes are intractable due to the presence of normalizing constants that require integration (sometimes stochastic) over a continuous domain. For an example, consider the inhomogeneous Poisson process with driving intensity $\lambda(\textbf{s})$. Its log-likelihood is written as
\begin{equation}
    \ell(\boldsymbol{\theta}; \mathcal{S})  = \sum_{i=1}^n\log\lambda(\textbf{s}_i) - \int_D\lambda(\bu)d\bu.
\end{equation}
With the exception of a few special cases, the integral for $\lambda(\textbf{s})$ cannot be found in closed form. In the case of a Cox process, likelihood based inference is complicated further since $\lambda(\textbf{s})$ is driven by an additional stochastic process. As such, evaluating the likelihood requires integrating over an infinite-dimensional random variable. 

Due to this likelihood intractability in many spatial point processes, a common approach is to maximize a simpler objective function that has the advantage of being computationally tractable. For spatial point processes, the Palm likelihood \citep{ogata1991maximum, tanaka2008parameter} is considered a type of pseudolikelihood where its construction arises from the Palm intensity. For example, for the inhomogeneous Poisson process, the log Palm likelihood is written
\begin{equation}\label{palm likelihood}
    \ell_p(\boldsymbol{\theta};\mathcal{S}) = \sum_{i\neq j}w_{ij}\log \lambda_p(\textbf{s}_i|\textbf{s}_j;\boldsymbol{\theta}) - \int_Dw_{uj}\lambda_p(\textbf{u}|\textbf{s}_j;\boldsymbol{\theta})d\textbf{u}
\end{equation}
where $w_{ij}$ is a nonnegative weight function, as is commonly seen in composite likelihoods \citep{varin2011overview}. Letting $w_{ij}=I\{||\bs_i-\bs_j||<R\}$ for some distance $R$, the score of (\ref{palm likelihood}) is an asymptotically unbiased estimating equation \citep{prokevsova2013asymptotic, baddeley2017local}. We approximate the integral term using numerical quadrature at evenly-spaced grid points $\bu_k$ with each grid cell having equal Lebesgue measure $|\bu|$.

The Palm likelihood is computationally advantageous because it is linear in the number of pairs of points, meaning that likelihood evaluation is $O(n^2)$ at worst. However, removing pairwise distances greater than $R$ can dramatically reduce this complexity for modest values of $R$. That being said, some computational burden still remains for likelihood evaluation. We present a discretized form of the log Palm likelihood in (\ref{palm likelihood}) for stationary, isotropic processes where we bin pairwise distances. For a stationary, isotropic process, the Palm likelihood is a function only of the distance between two points. Let $A^R=\{a_1,\dots, a_L\}$ denote a regularly spaced partition of $(0,R)$. Let $n_l = \sum_{i \neq j} \textbf{1}\{||\bs_i-\bs_j||\in a_l\}$ and $\tilde n_l = \sum_{i,k} \textbf{1}\{||\bs_i-\bu_k||\in a_l\}$. Letting $\lambda_p(a_l;\boldsymbol\theta)$ be the Palm intensity evaluated at the centroid of $a_l$, we propose a discretized approximation of the above likelihood where
\begin{equation}
    \ell_{p,A^R}(\boldsymbol\theta;\mathcal S) = \sum_{l=1}^Ln_l\log\lambda_p(a_l;\boldsymbol\theta)-\tilde n_l\lambda_p(a_l;\boldsymbol\theta)|\bu|,
    \label{eq:discLike}
\end{equation}
which has complexity $O(L)$, where $L$ is the cardinality of the partition of pairwise distances less than $R$. Assuming that each $a_l$ has equal measure, $\ell_{p,A^R}(\boldsymbol\theta;\mathcal S)$ can approximate $\ell_{p}(\boldsymbol\theta;\mathcal S)$ arbitrary well as $L\rightarrow\infty$. Thus, this discretized likelihood function is advantageous when a large $L$ is still smaller than the number of pairwise distances less than $R$. We illustrate in the online Appendix that the inference resulting from $\ell_{p,A^R}(\boldsymbol\theta;\mathcal S)$ is nearly identical to inference from $\ell_{p}(\boldsymbol\theta;\mathcal S)$.

\cite{tanaka2008parameter} first constructed the Palm likelihood for parameter estimation and model selection in Neyman-Scott models. More generally, this specification is suitable for many different processes which have simple, closed-form expressions and can be used for parameter estimation. Asymptotic properties for the maximum Palm likelihood estimator are established in \cite{prokevsova2013asymptotic} under the increasing-domain framework. 
Consider an increasing sequence of domains $\mathcal D_n$ where all $\mathcal D_n$ are convex sets and $\mathcal D_n\subseteq \mathcal D_{n+1}$ for all $n\in\mathbb N$. Denote $\mathcal S_n$ as the observed point pattern on $\mathcal D_n$. Letting 
\begin{equation}
    \hat{\boldsymbol\theta}_n = \arg\min_{\boldsymbol{\theta}} 
 \ell_p(\boldsymbol\theta;\mathcal{S}_n),
\end{equation}
$\hat{\boldsymbol\theta}_n$ is shown to be asymptotically normal with variance $G^{-1}(\boldsymbol{\theta}_0)=H^{-1}(\boldsymbol{\theta}_0)J(\boldsymbol{\theta}_0)H^{-1}(\boldsymbol{\theta}_0)$. The matrix $G^{-1}(\boldsymbol{\theta}_0)$ is known as the Godambe information \citep{godambe}, where $H(\boldsymbol{\theta}_0)=E[\frac{d^2}{d\boldsymbol\theta^2}\ell_{P}(\mathcal{S};\boldsymbol{\theta}_0)]$ and $J(\boldsymbol\theta_0)=Var(\frac{d}{d\boldsymbol{\theta}}\ell_{P}(\mathcal{S};\boldsymbol{\theta}_0))$.

While the Palm likelihood can be optimized directly for parameter estimation, when the mean and covariance parameters are separable it is often done in a two-step procedure in practice \citep{waagepetersen2009two}. For example, let the parameter vector be defined as $\boldsymbol{\theta}=\{\boldsymbol{\beta},\boldsymbol{\phi}\}$ where $\boldsymbol\beta$ is the vector of mean parameters and $\boldsymbol\phi$ is the vector of covariance parameters. The two-step procedure, similar to a profile likelihood approach, starts with estimating $\boldsymbol\beta$ by solving the score function
\begin{equation}
  \sum_{i=1}^n\frac{d\lambda(\bs_i)}{d\boldsymbol\beta}\frac{1}{\lambda(\bs_i)}-\int_\mathcal D \frac{d\lambda(\bu)}{d\boldsymbol\beta}d\bu=0.
\end{equation}
Then, the estimated $\hat{\boldsymbol\beta}$ is plugged into (\ref{palm likelihood}), which is then solved for $\boldsymbol{\phi}$. This procedure greatly simplifies the optimization process. In the constant mean case of the log-Gaussian Cox process, $\beta_0$ (the intercept term in $\boldsymbol{\beta}$) is biased for small values of $R$ when directly optimizing the Palm likelihood \citep{dvovrak2012moment}. This shortcoming is overcome with the two-step procedure as the intensity is being estimated directly as if in a homogeneous Poisson process. An important drawback of the two-step procedure is that uncertainty associated with underlying second-order properties is not taken directly into account when estimating standard errors of $\boldsymbol{\beta}$. This two-step procedure is employed in the popular \texttt{spatstat} package \citep{baddeley2005spatstat} in CRAN \citep{R}.

\subsection{Examples of Palm intensities}

The Palm intensity can be defined in closed form for a suite of different point processes exhibiting the spectrum of clustering to inhibition behaviors. This includes, for example, the log-Gaussian Cox process, the Neyman-Scott process, and the determinantal point process. Simulated realizations from these three point processes are shown in Figure \ref{fig:pp_examples}. Each of these point patterns exhibit different first and second order structure and yet can be defined by a closed form Palm intensity function. Below we define each of these point processes and specify their Palm intensities.

\subsubsection{Log-Gaussian Cox Process}

A log-Gaussian Cox process (LGCP) is an inhomogeneous Poisson process with intensity driven by a Gaussian process (GP). Let $W(\bs)$ denote a mean 0 spatial GP with covariance function parameterized by $\boldsymbol\phi$. 
For the moment, define the intensity function at location $\bs$ as $\lambda(\bs)=\exp\{\beta_0+W(\bs)\}$ where $\beta_0$ captures the average intensity for all $\bs$. Then, the parameter of interest is the vector $\boldsymbol\theta=\{\beta_0,\boldsymbol\phi\}$. 
The Palm intensity for the LGCP can be written 
\begin{equation}
    \lambda_p(\bs_i|\bs_j;\boldsymbol\theta) = \exp\{\beta_0 + \sigma^2/2 + \sigma^2\exp\{c(\bs_i,\bs_j;\boldsymbol\phi)\}\}
\end{equation}
where $c(\bs_i,\bs_j;\boldsymbol\phi) = \text{corr}(W(\bs_i),W(\bs_j))$, the spatial correlation of the GP between locations $\bs_i$ and $\bs_j$. 
This formulation can be extended to account for covariates by replacing the mean term $\beta_0$ with a regression such as $\textbf{x}(\bs_i)'\boldsymbol\beta$, where $\textbf{x}(\bs_i)$ is a vector of spatial covariates at location $\bs_i$. Under this specification, the parameter vector becomes $\boldsymbol\theta=\{\boldsymbol{\beta},\boldsymbol\phi\}$. 

\subsubsection{Neyman-Scott Process}\label{neyman_scott}

The class of Neyman-Scott processes exhibit clustering via parent and offspring points, while assuming stationarity and isotropy. Let $\boldsymbol\theta=\{\mu,\nu,\boldsymbol\phi\}$. To generate a realization from a Neyman-Scott process, first parent points are generated by a homogeneous Poisson process with intensity $\mu$. Then, for each generated parent point, the number and location of offspring points are generated.  
Specifically, the number of offspring points is randomly generated with mean $\nu$ and the distances are identically and independently distributed according to a specified density $f_{\boldsymbol\phi}(\boldsymbol r)$ where $\boldsymbol\phi$ denotes the parameters indexing the density function and $\boldsymbol r$ is the location vector of the offspring relative to the parent. 
The collection of offspring points define the realization of the Neyman-Scott process. 

Due to stationarity and isotropy, the Palm intensity function is only dependent upon the distance between points, $u$, giving $\lambda_p(\bs_i|\bs_j;\boldsymbol\theta)=\lambda_p(||\bs_i-\bs_j||)=\lambda_p(u)$. The Palm intensity for a Neyman-Scott process is defined through a relation with the K-function
\begin{equation}
    \lambda_p(u)=\lambda g(u)=\frac{\lambda}{2\pi  u}\frac{d}{du} K(u)
\end{equation}
\begin{equation}
    \lambda K(u)=\nu\{\mu\pi u^2+F_{\boldsymbol\phi} (u)\}
\end{equation}
where $K(u)$ is Ripley's K-function, $F_{\boldsymbol\phi}(u)$ is the distribution function of the distance between two points within the same cluster, and $\lambda=\mu\nu$ is the overall intensity.

A special case of the Neyman-Scott process is the Thomas process, where $\boldsymbol\phi = \sigma^2$ and
$f_{\boldsymbol\phi}(\boldsymbol r)$ is a bivariate normal density with mean zero and covariance $\sigma^2 \mathbf{I}_2$. For the Thomas process, the Palm intensity can be written
\begin{equation}
    \lambda_p(\bs_i|\bs_j;\boldsymbol{\theta})=\mu\nu +\frac{\nu}{4\pi\sigma^2}\exp\{-||\bs_i-\bs_j||^2/4\sigma^2\}
\end{equation}

\subsubsection{Determinantal Point Process}

Determinantal point processes (DPP) \citep{lavancier2015determinantal} are inhibition processes that have traditionally found applications in statistical physics and random-matrix theory. Recently, they have seen more use in spatial statistics, where the spatial locations of events reduce the probability of nearby events. Like cluster processes, inhibition processes are described through their second-order structure. Let $C(\bs_i,\bs_j;\boldsymbol \theta)$ be a semi-positive definite kernel defined on $\mathbb R^2\times\mathbb R^2$, typically assumed to be a covariance function between locations $\bs_i$ and $\bs_j$, and parameterized by $\boldsymbol \theta$. Then the $n^{th}$ product density is given by
\begin{equation}
    \lambda_n(\bs_1,\dots,\bs_n)=\det[\Sigma(\boldsymbol{\theta})].
\end{equation}
where $\Sigma(\boldsymbol{\theta})$ is the $n \times n$ covariance matrix constructed such that its $i,j$th element is $C(\bs_i,\bs_j;\boldsymbol\theta)$.
It follows from this construction that $\lambda(\bs)=C(\bs,\bs;\boldsymbol\theta)$ and $\lambda_2(\bs_i,\bs_j)=C(\bs_i,\bs_i;\boldsymbol\theta)C(\bs_j,\bs_j;\boldsymbol\theta)-C(\bs_i,\bs_j;\boldsymbol\theta)C(\bs_j,\bs_i;\boldsymbol\theta)$.
Then, the Palm intensity for the DPP can be written 
\begin{equation}
    \lambda_p(\bs_i|\bs_j;\boldsymbol\theta)=C(\bs_i,\bs_i;\boldsymbol\theta)-\frac{C(\bs_i,\bs_j;\boldsymbol\theta)C(\bs_j,\bs_i;\boldsymbol\theta)}{C(\bs_j,\bs_j;\boldsymbol\theta)}.
\end{equation}

The tractability of the moments of a DPP is desirable compared to Markov point processes, which have more complex moments requiring approximation. The drawback of these models is that they are slightly less flexible than Gibbs processes \citep{lavancier2015determinantal}, namely in that they cannot replicate the strong regularity of hard-core processes.

\begin{figure}
    \centering
    \includegraphics[width=1\linewidth]{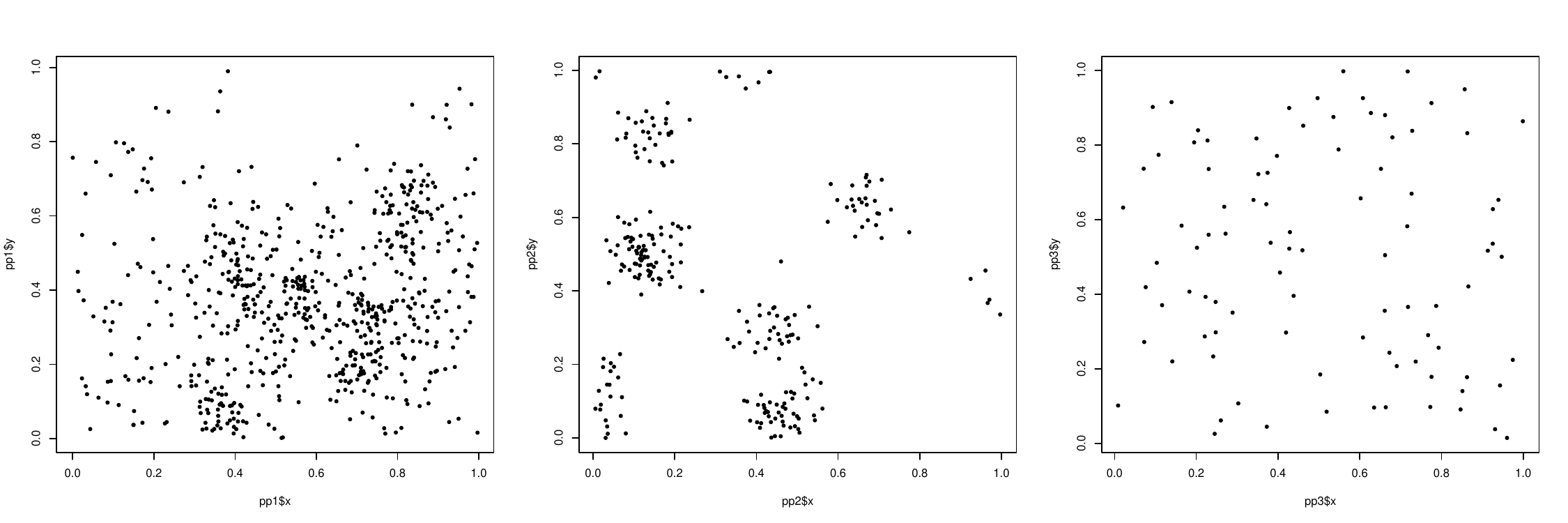}
    \caption{Realizations of (left) LGCP (middle) Thomas process and (right) Determinantal point process with exponential kernel}.
    \label{fig:pp_examples}
\end{figure}

The simplicity of the Palm intensity expressions for these processes exhibiting radically different behaviors (Figure \ref{fig:pp_examples}) is attractive for pseudolikelihood-based inference using (\ref{palm likelihood}). Palm intensities can also be obtained for Poisson processes, general Cox processes (such as shot-noise Cox processes \citep{moller2003shot}), and Gibbs processes \citep{coeurjolly2017tutorial}. For Poisson processes, the Palm intensity simplifies to $\lambda_p(\bs_i|\bs_j)=\lambda(\bs_i)$, rendering this procedure unnecessary. Cox processes inherit their moment properties from their latent stochastic processes, meaning the Palm intensity can be computed as long as their first two moments can be obtained. Unfortunately, the Palm intensity for Gibbs processes is not available in closed form \citep{coeurjolly2017tutorial}, necessitating the use of an approximation. Thus, the pseudolikelihood approach for Gibbs processes developed by \cite{baddeley2000practical} remains the most popular inferential technique.

\section{Bayesian Inference}

\subsection{Bayesian Composite Likelihoods}

Traditional Bayesian inference relies on deriving (or sampling from) the posterior distribution. 
Let $\pi(\boldsymbol{\theta)}$ define the prior distribution for the model parameters $\boldsymbol{\theta}$.
In our setting, the posterior distribution can be written
\begin{equation}
 \pi(\boldsymbol{\theta}|\mathcal{S})\propto L(\boldsymbol\theta;\mathcal S)\pi(\boldsymbol{\theta)}.   
\end{equation}
As mentioned above, the likelihood function $L(\boldsymbol{\theta};\mathcal S)$ is intractable in the case of many point processes, including the three discussed in Section 2. There is a rich literature on approaches for sampling from posteriors with intractable likelihoods, such as Variational Bayes \citep{tran2017variational} and Approximate Bayesian Computation (ABC) \citep{sunnaaker2013approximate}. Here, we focus on the composite likelihood \citep{pauli2011bayesian} and more broadly, generalized Bayesian inference \citep{bissiri2016general}. Note that the Palm likelihood is a special case of composite likelihood. We replace the full likelihood $L(\boldsymbol{\theta};\mathcal S)$ with the Palm likelihood $L_P(\boldsymbol{\theta};\mathcal S)$ (or its discretized approximation $L_{p,A^R}(\boldsymbol\theta;\mathcal S$) to derive the Palm posterior
\begin{equation}
    \pi_{P}(\boldsymbol{\theta}|\mathcal{S})\propto L_{P}(\boldsymbol{\theta};\mathcal S)\pi(\boldsymbol{\theta}).
\end{equation}

\cite{chernozhukov2003mcmc} derived asymptotic properties for this posterior distribution assuming some regularity conditions. Let $\pi_{P,n}(\boldsymbol{\theta}|\mathcal{S}_n)$ be the posterior obtained considering observations on the window $\mathcal D_n$, an increasing sequence of domains (described in Section \ref{point process background}). As $n\rightarrow\infty$, the posterior mean accumulates mass around $\hat{\boldsymbol\theta}_n$, the maximizer of the objective function \citep{chernozhukov2003mcmc}. Further, they show that 
\begin{equation}
    ||\pi_{P,n}(\boldsymbol{\theta}|\mathcal{S}_n)-\pi_{P,\infty}(\boldsymbol\theta|\mathcal{S}_\infty)||_{TV}\xrightarrow[]{p}0
\end{equation}
as $n\rightarrow\infty$ where $||\cdot||_{TV}$ is the total variation norm and $\pi_{P,\infty}(\boldsymbol{\theta}|\mathcal{S})$ is a normal density with variance $H^{-1}(\boldsymbol{\theta}_0)$. Recall that $\hat{\boldsymbol\theta}_n$ is asymptotically Normal with variance $G^{-1}(\boldsymbol{\theta}_0)=H^{-1}(\boldsymbol{\theta}_0)J(\boldsymbol{\theta}_0)H^{-1}(\boldsymbol{\theta}_0)$. This implies that the asymptotic variance of the Palm posterior and the asymptotic variance of the Palm likelihood estimator are not equivalent. As such, in order to obtain appropriate posterior inference, defined by nominal coverage of the resulting credible intervals, a posterior adjustment is required \citep{monahan1992proper}. In the next section we propose two different posterior adjustments for Bayesian inference for spatial point patterns using the Palm likelihood.

\subsection{Posterior Calibration}

Bayesian inference using the Palm likelihood can be shown to result in under-coverage of the posterior distributions of the model parameters. Heuristically, the invalid frequentist coverage in the generalized Bayesian framework is the result of the information in the likelihood relative to the prior being misrepresented. Under a correctly specified model and likelihood (henceforth full likelihood), the likelihood principle gives that all of the information in the data of size $n$ is represented in the likelihood. In the case of the Palm likelihood, $n\choose 2$ pairs of points are considered independently, thus dramatically inflating the relative strength of the likelihood. 

We propose two adjustments to achieve asymptotically valid credible intervals, wherein the $(1-\alpha)$ credible intervals achieve a Type I error rate of $\alpha$. The first approach involves raising the Palm likelihood to a fractional power based on theoretical asymptotic behavior in order to down-weight the amount of information present. The second approach offers a direct calibration of the posterior using simulation-based coverage checks.  

\subsubsection{Adjustment 1 - learning rate adjustment of the likelihood}

The first approach aims to modify the likelihood by raising it to a power, or "learning rate," in order to appropriately calibrate the posterior samples. 
The resulting posterior is written 
\begin{equation}
    \pi_{P}(\boldsymbol{\theta}|\mathcal{S})\propto L_{P}(\mathcal{S};\boldsymbol{\theta})^\eta\pi(\boldsymbol{\theta}).
\end{equation}
This approach naturally appears in a generalized Bayesian framework \citep{bissiri2016general}. We present relevant asymptotic results for the Palm likelihood to justify a particular selection of the "learning rate," denoted $\eta$. Then, we propose a procedure for estimating $\eta$ that is computationally efficient.

The role of $\eta$ is to appropriately weight the likelihood relative to the prior, in terms of the amount of information that each provides. An appropriate adjustment aims to \textit{down-weight} the Palm likelihood using $\eta\in(0,1)$. A survey of approaches to select the learning rate in the more general setting is presented in \cite{wu2023comparison} while \cite{pauli2011bayesian} provide an overview specific to composite likelihoods. We adopt the so-called "first-order moment matching" approach \citep{geys1999pseudolikelihood}. This approach was termed the "magnitude adjustment" in \cite{ribatet2012bayesian}, who use the composite likelihood in the context of max-stable processes for extreme value analysis. Based on the asymptotic properties of the Palm likelihood \citep{prokevsova2013asymptotic}, the log-likelihood ratio statistic asymptotically converges to a weighted sum of chi-squared random variables
\begin{equation}
    -2\{\ell_P(\mathcal{S}|\boldsymbol{\theta}_0)-\ell_P(\mathcal{S}|\boldsymbol{\hat\theta})\}\rightarrow\sum_{i=1}^q \lambda_i \chi^2_1
\end{equation}
where $\lambda_i$ are the eigenvalues of $H(\boldsymbol\theta_0)^{-1}J(\boldsymbol\theta_0)$. Thus, if we let $\eta=q/tr(H(\boldsymbol\theta_0)^{-1}J(\boldsymbol\theta_0))$, then
\begin{equation}
    E[-2\{\eta\ell_P(\mathcal{S}|\boldsymbol{\theta}_0)-\eta\ell_P(\mathcal{S}|\boldsymbol{\hat\theta})\}]\rightarrow\eta\sum_{i=1}^q\lambda_i = q.
\end{equation}
Note that the full log-likelihood ratio statistic converges in distribution to a chi-squared distribution with $q$ degrees of freedom. Thus, multiplying by $\eta$ here results in the first moment of the log Palm likelihood ratio converging to the same value as the full log-likelihood ratio, theoretically giving us adequate coverage.

Implementing this adjustment requires that we estimate $H(\boldsymbol\theta_0)^{-1}J(\boldsymbol\theta_0)$. Estimates are difficult to directly obtain because the theoretical expressions contain third and fourth moments of the process. Conveniently, since $H(\boldsymbol\theta_0)^{-1}$ is the asymptotic variance of the unadjusted posterior, we can obtain an estimate $\hat H(\boldsymbol\theta_0)^{-1}$ by fitting the unadjusted model with $\eta=1$. In the Bayesian framework, we estimate $\hat H(\boldsymbol\theta_0)^{-1}$ taking the empirical covariance of the resulting posterior samples of $\boldsymbol\theta$. Then, an estimate $\hat J(\boldsymbol\theta_0)$ is obtained via parametric bootstrap using the posterior means initially obtained as the generating parameters. Specifically, a prespecified number $B$ of bootstrapped point patterns $\{\mathcal{S}_1,\dots,\mathcal{S}_B\}$ are generated and the score $\frac{d}{d\boldsymbol\theta}\ell_p(\boldsymbol\theta;\mathcal S)$ is calculated for each. The resulting estimator is computed as
\begin{equation}
    \hat J(\boldsymbol\theta_0) = \widehat{Var}\bigg(\frac{d}{d\boldsymbol\theta}\ell_p(\hat{\boldsymbol\theta};\mathcal S_i)\bigg)
\end{equation}
where $\hat{\boldsymbol\theta}$ is the posterior mean with which the bootstrapped point patterns are generated.

\subsubsection{Adjustment 2 - matrix adjustment of the posterior samples}

The second adjustment we develop calibrates the posterior distribution by scaling each of the posterior samples by a matrix $\boldsymbol\eta$. This approach extends the work of \cite{shaby2014open} who proposes scaling (after centering) the posterior draws of $\boldsymbol\theta|\mathcal{S}$ obtained via MCMC to achieve desired frequentist coverage. For the $j^{th}$ posterior sample and a scaling matrix $\boldsymbol\eta$, we produce a calibrated sample via
\begin{equation}\label{theta transform}
    \tilde{\boldsymbol{\theta}}^{(j)} = \hat{\boldsymbol{\theta}}+\boldsymbol{\eta}(\boldsymbol{\theta}^{(j)}-\hat{\boldsymbol{\theta}})
\end{equation}
where $\hat{\boldsymbol\theta}$ is the posterior mean.
\cite{shaby2014open} uses an asymptotic approach to justify the selection of $\boldsymbol\eta$, which he calls the "open-face sandwich adjustment matrix". 
\cite{ribatet2012bayesian} proposed a similar approach where a "curvature adjustment" to the posterior samples of the parameter vector $\boldsymbol{\theta}$ is applied within each iteration of MCMC and evaluated via the likelihood. Namely, within the Metropolis-Hastings algorithm, each proposal for $\boldsymbol\theta^{(j)}$ is transformed via (\ref{theta transform}) and then $\tilde{\boldsymbol\theta}^{(j)}$ is evaluated in the acceptance-rejection step.

Here, we propose an alternative approach to selecting $\boldsymbol\eta$ via simulation -- namely, a bootstrap approach. 
Our approach leverages the work of \cite{syring2019calibrating}, who propose a generalized posterior calibration (GPC) approach as a means to select the scalar $\eta$ (not a matrix  as we are now considering) used as a learning rate. As with the previous adjustment, the goal of GPC is to ensure sufficient coverage of credible intervals. To do so, \cite{syring2019calibrating} use a nonparametric bootstrap to create $B$ bootstraps datasets. Then, posterior samples are obtained for each of these new datasets with the estimated $\boldsymbol{\hat\theta}$ as the "truth". Coverage rates are calculated and, if coverage is unsatisfactory, a new $\eta$ value is tried and the process continues iteratively until convergence within some tolerance. Each $\eta$ proposal in this algorithm requires an evaluation process of complexity $O(B\cdot C(n))$ where $C(n)$ is the cost of posterior computation given $n$ data points. Posterior samples can be obtained in parallel across the simulated datasets for a given $\eta$, but each proposed value of $\eta$ has to be evaluated sequentially.

For spatial point processes, we modify the GPC algorithm in several ways. Firstly, nonparametric bootstrapping of a point pattern has been studied little in the literature. Therefore, we opt for a parametric bootstrap under the assumed model using $\boldsymbol{\hat\theta}$ as the generating parameter values. Secondly, the post-hoc adjustment negates the need in \cite{syring2019calibrating} to iteratively compute bootstrapped posteriors to search for $\eta$. Recall that the computational complexity of the original Palm likelihood is linear in the number of pairwise distances less than $R$. Unfortunately, the number of points in a point process is random, so when simulating bootstrapped datasets, there can be a large variance in the computational complexity of each bootstrapped posterior computation. The proposed discretized version of the log Palm likelihood $\ell_{p,A^R}(\boldsymbol\theta;\mathcal S)$ in (\ref{eq:discLike}) is linear in the cardinality of the partition, which is constant across bootstrapped samples, meaning the overall complexity of the adjustment is $O(BL)$. This allows for a greater control over the computational cost. Lastly, by considering a matrix $\boldsymbol{\eta}$ instead of a scalar learning rate, we allow for unique scalings for each parameter. This is of particular interest because the learning rate approach inflates the uncertainty of all parameters equally, while the information in the data about each parameter is likely to vary substantially, particularly between main effect parameters and those of the covariance function. This results in more accurate empirical coverage of our adjusted posterior distributions for each model parameter. Note that our proposed approach constructs a diagonal matrix $\boldsymbol\eta$ in contrast to the approaches of \cite{ribatet2012bayesian} and \cite{shaby2014open} who can have nonzero off-diagonals. Our matrix adjustment approach selects $\boldsymbol{\eta}$ via the modified GPC algorithm presented in Algorithm \ref{alg:modified_gpc}.

\begin{algorithm}
\renewcommand{\baselinestretch}{0.9}\normalsize
\caption{Modified GPC Algorithm}
\label{alg:modified_gpc}
\begin{algorithmic}[1]
\Require Observed point pattern $\mathcal{S}$, number of MCMC samples $M$, number of bootstrap samples $B$, coverage level $1-\alpha$
\State \textbf{Step 1: Initial MCMC}
\State Obtain samples $\{\boldsymbol\theta^{(1)},\dots,\boldsymbol\theta^{(M)}\}$ from posterior $\pi_P(\boldsymbol\theta|\mathcal{S})$ using MCMC
\State Calculate posterior mean $\hat{\boldsymbol\theta}$ 
\Statex
\State \textbf{Step 2: Parametric Bootstrap (parallelize)}
\State Simulate $B$ independent spatial point processes $\mathcal{S}_1, \dots, \mathcal{S}_{B}$ using $\hat{\boldsymbol\theta}$ as the true parameter values.
\State For $k=1, \dots, B$, obtain samples $\{\boldsymbol\theta^{(1)}_k,\dots,\boldsymbol\theta^{(M)}_k\}$ from posterior $\pi_P(\boldsymbol\theta|\mathcal{S}_k)$ using MCMC

\Statex
\State \textbf{Step 3: Selecting $\boldsymbol\eta$}

\For{$i=1$ to $q$}
\State Denote $(1-\alpha)$ credible region of $\theta_i$ from posterior $k$ as $R_{1-\alpha,k}(\theta_i)$
\State Compute:
\[
\hat{p}_i = \frac{1}{B} \sum_{k=1}^{B} I\{\hat{\theta}_i \in R_{1-\alpha,k}(\theta_i)\}
\]
\If{$\hat{p} \geq (1-\alpha)$}
    \State \Return $\eta_i=1$
\Else
    \State Let $\tilde{\theta}_i^{(j)} = \hat{\theta}_i + \eta_i (\theta_i^{(j)} - \hat{\theta}_i)$ and denote the resulting credible regions $R_{1-\alpha,k}(\tilde\theta_i)$
    \State Numerically find $\eta_i$ such that:
    \[
    \hat{p}_i = \frac{1}{B} \sum_{k=1}^{B} I\{\hat{\theta}_i \in R_{1-\alpha,k}(\tilde\theta_i)\} \geq 1-\alpha
    \]
\EndIf
\EndFor

\Statex
\State \textbf{Step 4: Calibrate original chain}

\State For $j=1,\dots,M$, rescale original chain $\{\boldsymbol\theta^{(1)},\dots,\boldsymbol\theta^{(M)}\}$
\[
\tilde{\boldsymbol\theta}^{(j)} = \hat{\boldsymbol\theta} + \boldsymbol{\eta}(\boldsymbol\theta^{(j)} - \hat{\boldsymbol\theta})
\]
where \[
\boldsymbol{\eta} = \mathrm{diag}(\eta_1, \dots, \eta_q)
\]
\end{algorithmic}
\renewcommand{\baselinestretch}{1.0}\normalsize
\end{algorithm}

\subsection{Empirical Prior}

Our last contribution is in the form of an empirical prior that mimics the two-step procedure previously mentioned. Consider a spatial point process where the first-order intensity in (\ref{first order intensity}) is a constant, $\lambda(\bs)=\lambda$, and the second-order intensity is a function of parameter $\boldsymbol\theta$. The \texttt{spatstat} package uses a two-step estimation procedure wherein $\lambda$ is first estimated and then this plugin estimate is used in the Palm likelihood to estimate $\boldsymbol\theta$.

We propose an empirical prior to mimic the two-step estimator. First, we reparameterize the Palm intensity as $\lambda_p(\bs_i|\bs_j;\boldsymbol{\theta})=\lambda g(\bs_i,\bs_j;\boldsymbol\phi)$. 
Note that the maximum likelihood estimator (MLE) for $\lambda$ in the homogeneous Poisson process is $\hat\lambda=N(\mathcal{D})/|\mathcal D|$ where $N(\mathcal{D})$ is the number of observed points. Thus, we construct an empirical prior for $\lambda$ such that
\begin{equation}
    \lambda \sim N(N(\mathcal{D})/|\mathcal D|,\sigma^2_\lambda)
\end{equation}
Importantly, using composition sampling we can recover the posterior distribution of other parameters that are a function of $\lambda$, but not explicitly estimated. As an example, in the LGCP, the posterior of $\beta_0$ can be obtained as a function of the posteriors of $\lambda$ and $\sigma^2$, i.e. $\beta_0=\log(\lambda)-\sigma^2/2$. We explore this empirical prior approach in Section \ref{simulation study}.

\section{Simulation Study} \label{simulation study}

We now present simulation studies for three processes to examine the efficacy of our approach. First, we simulate from and fit a log-Gaussian Cox process. In using the Palm likelihood, we consider two different distance thresholds, $R$, posterior adjustments 1 and 2, and the empirical prior. We then compare our results to those from a full likelihood Bayesian approach. The second and third simulation studies illustrate the utility of the Palm likelihood approach for fitting a Thomas process and a determinantal point process. All MCMC sampling is conducted using the Stan programming language \citep{carpenter2017stan} (which employs an HMC algorithm) executed in R Version 4.4.1. Stan conveniently provides the option for employing user-defined functions in place of log-likelihood functions; therefore, it is straightforward to set the log Palm likelihood for the data level in the target density. For each model and simulation, Stan samples 10,000 iterations with 1,000 samples of burn-in, resulting in 9,000 samples available for posterior inference. Point patterns are generated via the \texttt{spatstat} package. In carrying out these experiments, we used a Macbook Pro (Apple M3 Pro 12-core CPU, 36 GB of RAM) running macOS Sonoma 14.5. When parallelization was possible in the GPC algorithm, we opted to use 10 cores.

\subsection{Log-Gaussian Cox Process}

We consider three simulation studies with varying number of expected points per unit square and domain size: (I) $\mathcal{D}=[0,1]^2$, $E[N(\mathcal D)]/|\mathcal D|=300$, (II) $\mathcal{D}=[0,1]^2$, $E[N(\mathcal D)]/|\mathcal D|=1200$, and (III) $\mathcal{D}=[0,2]^2$, $E[N(\mathcal D)]/|\mathcal D|=300$. The first setting is used to test inferential properties, but the latter two are used only for testing computational efficiency, so we simulate 100 spatial point patterns from the first, and only 10 from the latter two. We define the latent GP as mean 0 with exponential covariance function parameterized by variance $\sigma^2=1$ and range $\phi=0.1$. Following \cite{diggle2013spatial}, when fitting the model we reparameterize the variance and range to $\log(\sigma^2)$ and $\log(\phi)$ such that they each have an unconstrained parameter space. Additionally, to facilitate the implementation of the empirical prior, we reparameterize the mean intensity and estimate $\lambda$ where $\lambda=\exp\{\beta_0+\sigma^2/2\}$. Posterior estimates of $\beta_0$ can be obtained using composition sampling.

For each of the simulated point patterns, we first fit the full likelihood model (FL), the Palm likelihood model (PL$_R$), and the empirical Palm likelihood model (PLE$_R$), where the subscript $R$ denotes the threshold distance $R \in \{0.2,,0.4\}$. For PLE$_R$, we use the empirical prior $\lambda\sim N(N(\mathcal D)/|\mathcal D|,1)$. For the remaining parameters, we assume $\log(\sigma^2)\sim N(0,3)$ and $\log(\phi)\sim N(\log(0.1),0.09)$ which, when transformed, corresponds to 95\% of the prior mass on $\phi$ being placed on the interval $(0.05,0.18)$.

For FL, we use a discrete approximation for numerical integration based on a set of $k$ quadrature points, $\bu_1, \dots, \bu_k$. Let $W_i$ denote the GP evaluated at quadrature point $\bu_i$ representing a subregion with equal area $|\bu|$. Additionally, let $n_i=|\mathcal{S}\cap \bu_i|$.
We can approximate the log-likelihood as
\begin{equation}
	\ell(\beta_0,W;\mathcal{S})\propto\sum_{i=1}^k  n_i(\beta_0+W_i) - \sum_{i=1}^k|\bu_i|\exp\{\beta_0+W_i\},
\end{equation}
which is computationally tractable. We let the number of quadrature points scale with the domain size such that each unit square is divided into a $20\times 20$ grid.

The primary appeal of our approach is the computational scalability, which we assess by total runtime and effective sample size per second and report in Table \ref{tab:computation_times} for all simulations. To conserve space, we only show the times for PLE. In the LGCP setting, we can see a dramatic improvement in total computation time, even when accounting for the posterior calibration steps. The GPC adjustment takes longer than the asymptotic adjustment, but still takes under 2 minutes on average. Increasing $R$ from 0.2 to 0.4 roughly doubles computation time, which is to be expected given likelihood evaluation is linear in the number of partitions $L$. The improvement is most noticeable in setting (III) when $\mathcal D=[0,2]^2$ and we have a $40\times 40$ grid on which to evaluate the full likelihood, which is estimated to take approximately 5-6 days on average.

\begin{table}[]
    \footnotesize
    \centering
    \begin{tabular}{lcclrrrrrr}
       & & & & \multicolumn{3}{c}{\textbf{Time (min)}} & \multicolumn{3}{c}{\textbf{ESS/second}} \\
        \cmidrule(lr){5-7} \cmidrule(lr){8-10}
    Process & $\mathcal{D}$ & $E[N(\mathcal D)]/|\mathcal{D}|$ & Model & Unadj. & Adj. 1 & Adj. 2 & Unadj. & Adj. 1 & Adj. 2\\
\hline
LGCP & $[0,1]^2$ & 300 & FL & 86.33 &  &  & 0.14 &  & \\
&  &  & PLE$_{0.2}$ & 0.06 & 0.22 & 0.97 & 1038.70 & 126.45 & 59.78\\
& & & PLE$_{0.4}$ & 0.10 & 0.37 & 1.91 & 609.22 & 94.25 & 33.11\\
 \cmidrule(lr){2-10}
& $[0,1]^2$ & 1200 & FL & 77.72 &  &  & 0.41 &  & \\
&  & & PLE$_{0.2}$ & 0.08 & 0.42 & 1.33 & 463.74 & 67.82 & 27.91\\
& & & PLE$_{0.4}$ & 0.16 & 0.78 & 2.81 & 259.20 & 40.65 & 14.71\\
\cmidrule(lr){2-10}
& $[0,2]^2$ & 300 & FL & 8328$^*$ &  &  &  &  & \\
& & & PLE$_{0.2}$ & 0.07 & 0.23 & 1.07 & 748.08 & 87.91 & 46.61\\
& & & PLE$_{0.4}$ & 0.14 & 0.45 & 2.14 & 380.05 & 45.29 & 24.04\\
\hline
Thomas & $[0,1]^2$ & 300 & PLE$_{0.2}$ & 0.04 & 0.14 & 0.76 & 1932.83 & 517.25 & 114.35 \\
\hline
DPP & $[0,2]^2$ & 75 & PLE$_{0.2}$ & 0.04 & 4.00 & 1.12 & 2564.39 & 18.42 & 99.99 \\
\end{tabular}
    \caption{Average total runtime and effective sample sizes for all simulation studies. Adj. 1 and Adj. 2 correspond to the total runtime/ESS of the intial posterior computation and the respective adjustment. ESS/second is computed with respect to $\log\sigma^2$, $\log\nu$, and $\log\alpha$ for the LGCP, Thomas process, and DPP simulations respectively. The time denoted marked * is an approximation based on obtaining 100 samples after 10 burn-in, and multiplying the total runtime by 100. For this reason, ESS/second is not reported for this setting.}
    \label{tab:computation_times}
\end{table}

To evaluate the estimation performance of each model in simulation setting (I), we compute the bias and RMSE of the posterior mean with respect to the parameters of interest, which are reported in Table \ref{lgcp_bias_rmse}. In general, the PL and PLE models have a lower bias for $\log\sigma^2$, but PLE is more biased for $\log \phi$ and PL is more biased for $\beta_0$. RMSE provides a better comparison in that the FL performs as well or better than the other models, presumably because of the efficiency of using the full likelihood. There is not clear support for one value of $R$ over the other in terms of inference. However, this does suggest that $R=0.2$ is adequate relative to $R=0.4$, and thus more desirable given the improved computation times.

\begin{table}[]
    \centering
    \begin{tabular}{lcccccc}
        & \multicolumn{3}{c}{\textbf{Bias}} & \multicolumn{3}{c}{\textbf{RMSE}} \\
        \cmidrule(lr){2-4} \cmidrule(lr){5-7}
        Model & $\beta_0$ & $\log\sigma^2$ & $\log\phi$ & $\beta_0$ & $\log\sigma^2$ & $\log\phi$ \\
        \hline
FL & 0.05 & -0.27 & 0.14 & 0.25 & 0.41 & 0.20\\
PLE$_{0.2}$ & -0.02 & -0.13 & -0.21 & 0.29 & 0.39 & 0.39\\
PL$_{0.2}$ & -0.23 & -0.05 & 0.07 & 0.53 & 0.43 & 0.34\\
PLE$_{0.4}$ & -0.06 & -0.06 & -0.29 & 0.31 & 0.39 & 0.44\\
PL$_{0.4}$ & -0.19 & -0.05 & 0.03 & 0.45 & 0.42 & 0.42\\
    \end{tabular}
        \caption{Bias and RMSE of posterior mean for the LGCP across models. 
        \label{lgcp_bias_rmse}}
\end{table}

Finally, we examine the coverage properties of the credible intervals from each model, including the two calibration approaches. Table \ref{tab:lgcp_cov} shows the empirical coverage rates of FL along with the unadjusted and adjusted versions of PL and PLE. In general, the adjusted posteriors tend to be more diffuse than those from FL model. This is to be expected as the Palm likelihood is a pseudolikelihood that loses some information when compared to the full likelihood, as shown previously with the inflated RMSEs. Adjustment 1 significantly over-covers the nominal coverage level, whereas adjustment 2 performs comparably to FL empirical coverage rates. The dramatic undercovering of the unadjusted posteriors further illustrates why these adjustments are necessary. Though adjustment 1 has sound theoretical justification, in practice it is difficult to estimate the covariance matrix of the maximum Palm likelihood estimator and that difficulty translates to too small of an $\eta$, giving larger credible intervals, as can be seen in the average credible interval lengths given in parentheses.

\begin{table}[ht]
    \centering
    \resizebox{\textwidth}{!}{%
    \begin{tabular}{lccccccccc}
    
        & \multicolumn{3}{c}{$\beta_0$} & \multicolumn{3}{c}{$\log\sigma^2$} & \multicolumn{3}{c}{$\log\phi$} \\
        \cmidrule(lr){2-4} \cmidrule(lr){5-7} \cmidrule(lr){8-10}
        Model & Unadj. & Adj. 1 & Adj.2 & Unadj. & Adj. 1 & Adj.2 & Unadj. & Adj. 1 & Adj.2 \\ \hline
FL & 0.97 (0.96) &  &  & 0.88 (1.11) &  &  & 0.96 (0.90) &  & \\
PLE$_{0.2}$ & 0.16 (0.11) & 0.85 (0.79) & 0.93 (0.98) & 0.21 (0.28) & 0.91 (6.46) & 0.87 (1.21) & 0.2 (0.29) & 1.00 (1.16) & 0.87 (1.35)\\
PL$_{0.2}$ & 0.26 (0.42) & 1.00 (1.82) & 0.89 (2.24) & 0.21 (0.35) & 1.00 (7.14) & 0.93 (1.54) & 0.67 (0.71) & 1.00 (1.17) & 0.97 (1.64)\\

PLE$_{0.4}$ & 0.14 (0.10) & 0.91 (0.88) & 0.87 (1.02) & 0.24 (0.25) & 1.00 (6.88) & \textbf{0.93} (1.45) & 0.12 (0.23) & 1.00 (1.16) & 0.81 (1.37)\\

PL$_{0.4}$ & 0.11 (0.13) & 1.00 (1.69) & 0.88 (1.57) & 0.19 (0.25) & 1.00 (7.22) & 0.97 (1.79) & 0.25 (0.36) & 1.00 (1.17) & 0.97 (1.85)\\
\end{tabular}
}
    \caption{Empirical coverage rates of the 95\% credible intervals for the LGCP parameters across each model and posterior adjustment. Average credible interval length is reported in parentheses.}
    \label{tab:lgcp_cov}
\end{table}

\subsection{Thomas Process}

The second simulation study uses the Thomas process, described in Section \ref{neyman_scott}, where the offspring points are distributed according to a bivariate Gaussian distribution. We simulate 100 point patterns from a Thomas process with generating parameters $\mu=10,\nu=30,\sigma^2=0.0025$ using the \texttt{spatstat} package. Similar to the LGCP, we reparameterize each parameter to the log scale and again model the $\lambda$ parameter where we can obtain the posterior distribution for $\nu$ via composition sampling with $\lambda=\mu\nu$. Uninformative normal priors are used for all three parameters. We choose $R=0.2$ because of the balance of computational and statistical performance shown in the LGCP study for this threshold. In Table \ref{tab:computation_times}, we see that the entire computation time for adjustment 2, the most expensive, comes in under 1 minute.

Table \ref{tab:thomas_bias_rmse} shows the bias and RMSE of the posterior mean parameter estimates across the 100 simulated datasets. Here, the non-empirical version performs better than the empirical version with respect to bias and RMSE across all parameters. Table \ref{tab:thomas_cov} shows empirical coverage rates for 95\% credible intervals under each adjustment. Again, adjustment 1 is far too conservative, whereas adjustment 2 is close to the nominal coverage rate. Based on these simulations we recommend the Thomas process be fitted without the empirical prior and utilizing the GPC adjustment.

\begin{table}[ht]
\centering
\centering
\begin{tabular}[t]{lcccccc}

\multicolumn{1}{c}{ } & \multicolumn{3}{c}{Bias} & \multicolumn{3}{c}{RMSE} \\
\cmidrule(l{3pt}r{3pt}){2-4} \cmidrule(l{3pt}r{3pt}){5-7}
Model & $\log\mu$ & $\log\nu$ & $\log\sigma^2$ & $\log\mu$ & $\log\nu$ & $\log\sigma^2$\\
\midrule
PLE$_{0.2}$ & 0.14 & -0.16 & -0.13 & 0.41 & 0.41 & 0.36\\
\hline
PL$_{0.2}$ & -0.07 & -0.01 & 0.00 & 0.39 & 0.34 & 0.29 \\

\end{tabular}
\caption{Bias and RMSE of posterior mean for the Thomas process across models.}
\label{tab:thomas_bias_rmse}
\end{table}

\begin{table}[ht]
    \centering
    \resizebox{\textwidth}{!}{%
    \begin{tabular}{lccccccccc}
    
        & \multicolumn{3}{c}{$\log\mu$} & \multicolumn{3}{c}{$\log\nu$} & \multicolumn{3}{c}{$\log\sigma^2$} \\
        \cmidrule(lr){2-4} \cmidrule(lr){5-7} \cmidrule(lr){8-10}
        Model & Unadj. & Adj. 1 & Adj.2 & Unadj. & Adj. 1 & Adj.2 & Unadj. & Adj. 1 & Adj.2 \\ \hline
PLE$_{0.2}$ & 0.18 (0.20) & 1.00 (15.64) & 0.90 (1.51) & 0.25 (0.20) & 1.00 (15.64) & 0.90 (1.44) & 0.32 (0.27) & 1.00 (11.79) & 0.89 (1.39)\\
\hline
PL$_{0.2}$ & 0.52 (0.61) & 1.00 (16.14) & 0.99 (2.26) & 0.45 (0.48) & 1.00 (15.99) & 1.00 (1.83) & 0.55 (0.49) & 1.00 (11.76) & 0.96 (3.80)\\
\hline

\end{tabular}
}
    \caption{Empirical coverage rates of the 95\% credible intervals for the Thomas process parameters across each model and posterior adjustment. Average credible interval length is reported in parentheses.}
    \label{tab:thomas_cov}
\end{table}

\subsection{Determinantal Point Process}

The final simulation study simulates from and fits a determinantal point process with a Gaussian kernel, such that $C(\bs_i,\bs_j;\lambda,\alpha)=\lambda\exp\{-(||\bs_i-\bs_j||/\alpha)^2\}$. We let $\alpha=0.05$, $\lambda=75$, and $\mathcal{D}=[0,2]^2$, such that $E[N(\mathcal D)]=300$. We simulate 100 point patterns from this process. In Table \ref{tab:computation_times}, we see a different result from the previous simulations in that adjustment 1 actually takes longer than adjustment 2. Importantly, simulating from a DPP is more expensive than the other two point processes considered, and to estimate the covariance matrix for adjustment 1 requires the simulation of 1000 point patterns. Therefore, the majority of the computation cost of calibrating the posterior comes from simulation.

Table \ref{tab:dpp_cov} shows that PLE$_{0.2}$ performs slightly better in point estimation, but slightly undercovers in comparison to the non-empirical approach. Under PLE$_{0.2}$, adjustment 1 results in too narrow of a credible interval for $\lambda$. In this case, the chosen $\eta$ was too large, and did not dramatically reduce the information in the likelihood, which may be a result of the challenges again in estimating the asymptotic covariance of the maximum Palm likelihood estimator.

\begin{table}[ht]
\centering
\centering
\begin{tabular}[t]{lcccc}

\multicolumn{1}{c}{ } & \multicolumn{2}{c}{Bias} & \multicolumn{2}{c}{RMSE} \\
\cmidrule(l{3pt}r{3pt}){2-3} \cmidrule(l{3pt}r{3pt}){4-5}
Model & $\lambda$ & $\log\alpha$ & $\lambda$ & $\log\alpha$\\
\midrule
PLE$_{0.2}$ & -0.14 & -0.01 & 3.91 & 0.12\\
PL$_{0.2}$ & -0.26 & -0.01 & 4.23 & 0.12\\

\end{tabular}
\caption{Bias and RMSE of posterior mean for the DPP across models.}
\label{tab:dpp_bias_rmse}
\end{table}

\begin{table}[ht]
    \centering
    \resizebox{\textwidth}{!}{%
    \begin{tabular}{lcccccc}
    
        & \multicolumn{3}{c}{$\lambda$} & \multicolumn{3}{c}{$\log\alpha$}  \\
        \cmidrule(lr){2-4} \cmidrule(lr){5-7} 
        Model & Unadj. & Adj. 1 & Adj.2 & Unadj. & Adj. 1 & Adj.2 \\ \hline
PLE$_{0.2}$ & 0.34 (3.30) & 0.40 (3.77) & 0.94 (14.18) & 0.85 (0.35) & 1.00 (1.69) & 0.92 (0.43)\\
\hline
PL$_{0.2}$ & 0.51 (6.14) & 0.98 (21.01) & 0.95 (15.98) & 0.88 (0.37) & 1.00 (5.14) & 0.95 (0.46)\\
\hline

\end{tabular}
}
    \caption{Empirical coverage rates of the 95\% credible intervals for the DPP parameters across each model and posterior adjustment. Average credible interval length is reported in parentheses.}
    \label{tab:dpp_cov}
\end{table}

\section{Data Analysis}

\begin{figure}
    \centering
    \includegraphics[width=1\linewidth]{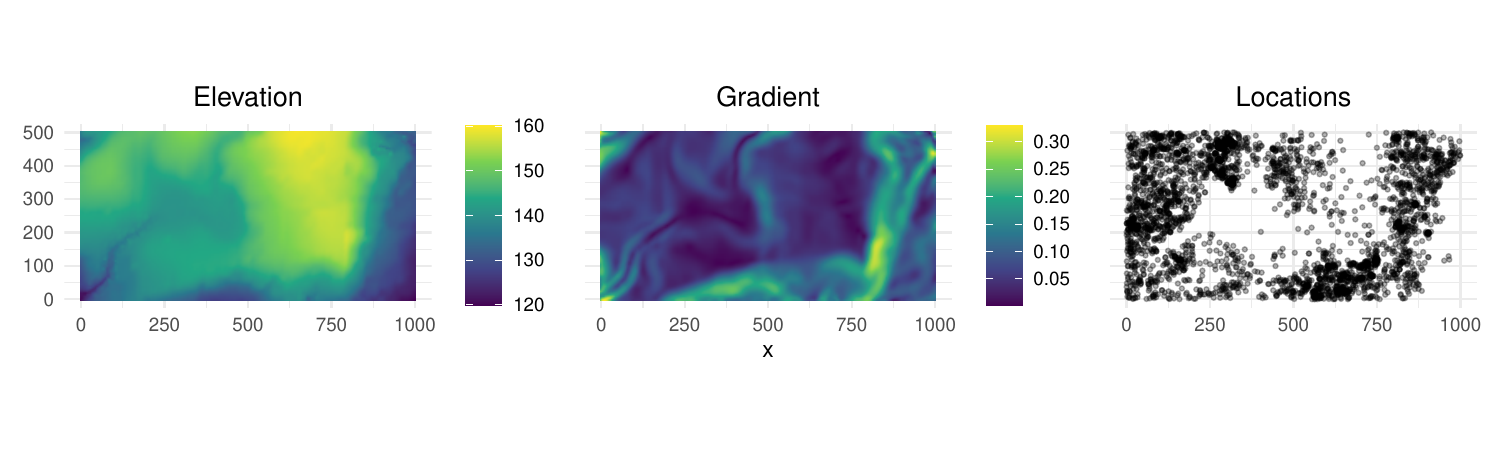}
    \vspace*{-2cm}
    \caption{Maps of spatial covariates, elevation and gradient, and the locations of \textit{Beilschmiedia pendula Lauraceae} trees.}
    \label{fig:bei}
\end{figure}

We apply our approach to the \texttt{bei} dataset found in \texttt{spatstat}. This dataset consists of 3605 \textit{Beilschmiedia pendula Lauraceae} trees located in the rainforest of Barro Colorado island within a 1000m $\times$ 500m spatial domain. The locations are shown spatially in Figure \ref{fig:bei}.  This dataset has been analyzed previously in the literature using other approaches \citep{moller2007modern,flagg2023integrated}.  
Spatial point patterns are often studied in forestry to detect important drivers of the spatial distribution of a particular species across a spatial region \citep{stoyan2000recent,comas2007modelling}. In our analysis we consider two possible explanatory variables, elevation and gradient, which are assumed to be possible drivers of the distribution of this tree species. Elevation and gradient are shown spatially in Figure \ref{fig:bei}.

We proceed by fitting an LGCP with covariates in the intensity function. Let $\mathcal{S}=\{\bs_1,\dots,\bs_n\}$ be the locations of trees with $n=3605$, $W(\bs_i)$ be the zero-mean latent Gaussian process at point $\bs_i$, and $\textbf{x}(\bs_i)$ be the vector of covariates (including an intercept term) at location $\bs_i$ evaluated at 20,000 grid locations. We assume $\mathcal{S}$ comes from an LGCP with log-intensity

\begin{equation}
    \log \lambda(\bs)=\textbf{x}(\bs)'\boldsymbol\beta + W(\bs)
\end{equation}
\begin{equation}
    \textbf{W}\sim GP(\textbf{0},\boldsymbol\Sigma)
\end{equation}
where the covariance matrix, $\boldsymbol{\Sigma}$, is defined by an exponential covariance function such that $Cov(W(\bs_i),W(\bs_j))=\sigma^2\exp\{||\bs_i-\bs_j||/\phi\}$. The parameters of interest are $\boldsymbol\beta=\{\beta_0,\beta_1,\beta_2\}$, $\sigma^2$, and $\phi$. For this model, the Palm intensity is written
\begin{equation}
    \lambda_p(\bs_i|\bs_j;\boldsymbol\theta)=\exp\{\textbf{x}(\bs_i)'\boldsymbol\beta+\sigma^2/2+ \sigma^2\exp\{||\bs_i-\bs_j||/\phi\}\}
\end{equation}
and $w_{ij}=I\{||\bs_i-\bs_j||\leq R\}$. 

To complete the Bayesian hierarchical model framework, we assign prior distributions to all model parameters. The regression coefficients $\beta_0$, $\beta_1$, and $\beta_2$ are each assigned independent, uninformative $N(0,1000)$ priors. Additionally, $\log(\sigma^2)\sim N(0,3)$, and $\log(\phi)\sim N(\log(100),0.25)$. We set $R=200$, which is $20\%$ of the long side of the rectangular region.

We compare our results to \cite{flagg2023integrated} and \cite{moller2007modern} who use INLA and a fully Bayesian approach for model inference, respectively. Each of these analyses assume different prior distributions on the parameters, but all are relatively uninformative priors. The only difference in model specification is that \cite{flagg2023integrated} use a Matérn covariance function with smoothness parameter $\nu=1$, whereas $\nu=1/2$ (the exponential covariance function) here and in \cite{moller2007modern}. 

\begin{table}[]
\scriptsize
    \centering
    \begin{tabular}{lccccc}
         & $\beta_0$ & $\beta_1$ & $\beta_2$ & $\sigma^2$ & $\phi$ \\ \hline
        MW & N/A & $0.06 (0.02,0.10)$ & $8.76(6.03,11.37)$ & $2.59(2.07,3.42)$ & $42.5(32.10,56.45)$ \\
        FH & $-10.9(-14.2,-7.63)$ & $0.03(0.01,0.05)$ & $4.44(2.42,6.46)$ & $1.69(1.10,2.66)$ & $176(137,229)$\\
        $PL_{200}$ & $-8.98(-14.61,-3.52)$ & $0.02(-0.02,0.06)$ & $4.36(0.99,7.77)$ & $1.29(0.90,1.80)$ & $62.10(34.27,107.24)$\\
    \end{tabular}
    \caption{Posterior means along with 95\% credible intervals in parentheses for parameters in LGCP model of \textit{Beilschmiedia pendula Lauraceae}.}
    \label{tab:bei}
\end{table}

The initial sampling before adjustment took 5.64 minutes. We parallelized over 5 cores, and the ensuing GPC calibration took approximately 10 hours. Due to sharing computing resources across cores, each bootstrapped MCMC took approximately 20 minutes, except for one single sampler that took closer to 200 minutes, most likely due to Stan's sampling procedure. Without this single aberration, total runtime would have been closer to 7 hours. The results shown in Table \ref{tab:bei} are labeled as MW for \cite{moller2007modern}, FH for \cite{flagg2023integrated}, and PL$_{200}$ for our approach. Posterior means and 95\% credible intervals are reported. Note that the intercept $\beta_0$ was not reported in \cite{moller2007modern}. Overall, we obtain similar results to the other analyses conducted with different methodologies. The most notable difference is that our approach has greater uncertainty, particularly for the $\beta_1$ and $\beta_2$ posteriors. This is to be expected based on our previous discussions. Although $\beta_1$ may not be deemed "significant" at the 95\% level, the credible region still provides evidence of a positive effect from elevation on tree intensity. Another notable difference is that the FH estimate of $\phi$ is larger than the other approaches, which we attribute to the difference in smoothness parameters used.

\section{Discussion}

In this paper, we have developed an approach for spatial point pattern analysis using the Palm likelihood in a generalized Bayesian framework. Motivated by the intractability of the full likelihood for many different point processes, we explored the efficacy of Palm likelihoods for Bayesian parametric inference. It was shown through simulation studies that significant computational gains are achieved with this method. However, a drawback is that the immediate posterior uncertainty when using the Palm likelihood does not provide adequate frequentist coverage. In order to remedy this shortcoming, we propose two different adjustments to calibrate the posterior. One modifies the "weight" of the likelihood using asymptotically justified estimates, while the other directly modifies the posterior samples using the modified GPC algorithm. Given the results of the simulation studies, we recommend adjustment 2 (GPC), as adjustment 1 is generally too conservative. Adjustment 1 does have a slight computational advantage in the LGCP and Thomas process settings, but not enough of an advantage to justify its conservatism. Additionally, we notice some bias in estimating the mean intensity in the LGCP, which we remedy by proposing an empirical prior to mimic traditional two-step estimation. The other processes perform comparably or better without the empirical prior.

The primary limitation of our work is that the inherent uncertainty (after calibration) when using the Palm likelihood is greater than under the fully Bayesian approach. This inefficiency can be attributed to the general loss of information that comes with using pseudolikelihoods in any context. However, given that our method is fast and easy to implement across a number of point process models, it is particularly useful for model selection and comparison in preliminary analysis. Additionally, there are scenarios such as in the LGCP simulation study, where likelihood evaluation on even a modest $40\times40$ grid is infeasible. The Palm likelihood offers an alternative in cases where the domain is too large for full likelihood inference. 

This paper is a first step into applying the generalized Bayesian framework to point pattern analysis. Future work includes comparing the performance of composite likelihoods \citep{guan2006composite} to the Palm likelihood in a Bayesian context, or even more general optimal estimating equations \citep{moller2017some}. Similarly, under the paradigm of \cite{bissiri2016general}, one could update prior beliefs under any loss function, such as the minimum contrast estimator (\ref{mce}). Additionally, the effects of the distance threshold $R$ on the resulting posterior merit a more in-depth study. Although the determinantal point process characterizes inhibition, Markov point processes are considered to be more flexible and are more commonly applied in the literature. Future work looks to extend our method using the pseudolikelihood for Gibbs processes \citep{baddeley2000practical}. Finally, asymptotic guarantees for the GPC algorithm have not yet been established for several reasons discussed in \cite{syring2019calibrating}. Developing such a result would be a significant contribution. Alternatively, different approaches for calibrating general posteriors have been developed \citep{wu2023comparison} and could be tested in this context.

\addtolength{\textheight}{-.2in}%

\section{Disclosure statement}\label{disclosure-statement}

The authors have no conflicts of interest to declare.

\section{Data Availability Statement}\label{data-availability-statement}

Data is available in the \texttt{spatstat} R package.

\phantomsection\label{supplementary-material}
\bigskip

\begin{center}

{\large\bf SUPPLEMENTARY MATERIAL}

\end{center}

\begin{description}
\item[APPENDIX:] An online appendix illustrating the suitability of the discretized log Palm likelihood approximation is included.
\item[CODE:]
The code to reproduce results in this manuscript can be found in the GitHub repository: https://github.com/kecollins/bayesian-palm

\end{description}

\bibliography{bibliography}

\newpage
\begin{center}
	{\LARGE\bf Appendix for ``Efficient Bayesian Inference for Spatial Point Patterns Using the Palm Likelihood"}
\end{center}
\medskip

We provide a brief comparison of posteriors obtained using the log Palm likelihood
\begin{equation}
	\ell_p(\boldsymbol{\theta};\mathcal{S}) = \sum_{i\neq j}w_{ij}\log \lambda_p(\textbf{s}_i|\textbf{s}_j;\boldsymbol{\theta}) - \int_Dw_{uj}\lambda_p(\textbf{u}|\textbf{s}_j;\boldsymbol{\theta})d\textbf{u}
\end{equation}
and the discretized approximation of the log Palm likelihood
\begin{equation}
	\ell_{p,A^R}(\boldsymbol\theta;\mathcal S) = \sum_{l=1}^Ln_l\log\lambda_p(a_l;\boldsymbol\theta)-\tilde n_l\lambda_p(a_l;\boldsymbol\theta)|\bu|,
\end{equation}
both of which are defined in Section 2 of the manuscript. We simulate 10 datasets from a log-Gaussian Cox process on the domain $\mathcal D=[0,1]^2$, as described in Section 4 of the manuscript, with $E[N(\mathcal{D})]=300$, $\sigma^2=1$, and $\phi=0.1$. For each dataset, we obtain the unadjusted posterior with an empirical prior using both $\ell_p(\boldsymbol\theta;\mathcal S)$ and $\ell_{p,A^R}(\boldsymbol\theta;\mathcal S)$ for $R=0.2$. We try three different values of $L=\{500,1000,1500\}$ to see how the resulting posterior is affected by the discretization scheme.

In Table \ref{tab:disc_comp}, we report the bias between $\ell_p(\boldsymbol\theta;\mathcal S)$ and $\ell_{p,A^R}(\boldsymbol\theta;\mathcal S)$ for the posterior mean, 0.025 posterior quantile, and 0.975 posterior quantile of each parameter. We compute this for each value of $L$. As we can see, the approximation error is negligible, and does tend to improve slightly with larger values of $L$, though the error is so small that this is not always a consistent pattern.

\begin{table}[H]
	\centering
	\begin{tabular}{lccccccccc}
		& \multicolumn{3}{c}{Mean} & \multicolumn{3}{c}{0.025 Quantile} & \multicolumn{3}{c}{0.975 Quantile } \\
		\cmidrule(lr){2-4} \cmidrule(lr){5-7} \cmidrule(lr){8-10}
		$L$ & $\beta_0$ & $\log\sigma^2$ & $\log\phi$ & $\beta_0$ & $\log\sigma^2$ & $\log\phi$ & $\beta_0$ & $\log\sigma^2$ & $\log\phi$ \\ \hline
		500 & 0.0005 & -0.0010 & 0.0014 & 0.0000 & -0.0023 & 0.0017 & 0.0009 & -0.0006 & 0.0045\\
		1000 & 0.0002 & -0.0005 & 0.0005 & -0.0008 & -0.0018 & -0.0010 & 0.0007 & 0.0020 & 0.0020\\
		1500 & 0.0003 & -0.0008 & 0.0009 & 0.0002 & -0.0017 & 0.0001 & 0.0005 & -0.0002 & 0.0022\\
		
	\end{tabular}
	\caption{Bias for posterior means, 0.025 posterior quantiles, and 0.975 posterior quantiles of each parameter under the three tested values of $L$.}
	\label{tab:disc_comp}
\end{table}

\end{document}